\documentclass[12pt]{iopart}
\usepackage{graphicx}
\usepackage{dcolumn}
\usepackage{bm}
\usepackage{epsf}

\begin{document}

\title{Multiphoton Bloch-Siegert shifts and level-splittings in a
  three-level system}
\author{P L Hagelstein$^1$, I U Chaudhary$^2$}

\address{$^1$ Research Laboratory of Electronics, 
Massachusetts Institute of Technology, 
Cambridge, MA 02139,USA}
\ead{plh@mit.edu}

\address{$^2$ Research Laboratory of Electronics, 
Massachusetts Institute of Technology, 
Cambridge, MA 02139,USA}
\ead{irfanc@mit.edu}

\begin{abstract}
In previous work we studied the spin-boson model in the multiphoton regime, 
using a rotation that provides a separation between terms that contribute most
of the level energies away from resonance, and terms responsible for the level
splittings at the anticrossing.
Here, we consider a generalization of the spin-boson model consisting of a three-level
system coupled to an oscillator.
We construct a similar rotation and apply it to the more complicated model.
We find that the rotation provides a useful approximation to the energy levels
in the multiphoton region of the new problem.
We find that good results can be obtained for the level splittings at the
anticrossings for resonances involving the lower two levels
in regions away from accidental or low-order resonances of the
upper two levels.   
\end{abstract}

\submitto{\jpb}

\maketitle

\section{Introduction}
\label{sec:intro}

   In early studies of spin dynamics in magnetic fields, Bloch and Siegert
considered the basic problem of the two-level system with a sinusoidal
perturbation \cite{BlochSiegert}.  
The perturbation increases the level separation (the Bloch-Siegert shift),
and energy can be exchanged with the two-level system when the transition
energy matches an odd multiple of $\hbar \omega_0$ (Bloch-Siegert resonances), 
where $\omega_0$ is the oscillatory frequency.
These effects have been studied with a dynamical Hamiltonian \cite{BlochSiegert,Shirley} (the Rabi Hamiltonian),
and also with a static Hamiltonian \cite{Cohen} (the spin-boson Hamiltonian) in which the two-level system
and oscillator are modeled as coupled quantum systems.
These models have been of interest over the years for applications to physical systems,
including spin problems \cite{BlochSiegert,Pegg,AhmadBullough} and atoms in strong
electromagnetic fields 
\cite{Shirley,HattoriKobayashi,Forre,OstrovskyHorsdal};  
and also as standard model problems on which new approximation methods can be tested 
\cite{Graham,Ciblis}
.
We have been interested in such problems in order to better understand
energy exchange between quantum systems with highly mismatched characteristic
energies.

In this work, we address a generalization of the spin-boson problem in which a three-level
system is coupled to an oscillator.
Our interest in this problem is focused primarily on the multiphoton regime, in which the
characteristic energy of the oscillator is small compared to the available transition
energies of the three-level system.
While there exists a modest literature on coupled oscillator and
three-level models \cite{WangHoChu,DAndrea,MatisovMazets,LiuLinZhu,KlimovSainz}, 
there does not appear to be available previous work relevant to the multiphoton region.
The rotating wave approximation has been applied with success near
low-order resonances \cite{RadmoreKnight,LiLin,Cardimona,WuYang,KlimovSoto,Kamli,AbdelWahab},
but one would not expect this approach to be helpful in the multiphoton region.

The replacement of the two-level system of the spin-boson problem by a three-level system in
the generalization of the model considered here very much complicates the problem.
The convenient mathematical machinery of the spin operators and Pauli matrices 
available for the spin-boson problem is now replaced by the less convenient SU(3)
operators and Gell-Mann matrices \cite{Georgi}, and the diagonalization of the three-level system
now involves a cubic characteristic equation.
There can occur in addition level crossing effects in the three-level system
in which the middle level is pushed through one of the other levels. 
Faced with such difficulties, it is no surprise that the three-level version of the
model has not received comparable attention as a workhorse problem 
in comparison with the simpler spin-boson problem.

Here, we propose to use the more complicated coupled oscillator and three-level model to
test an approximation scheme that appears to have worked well in the spin-boson problem \cite{HagelsteinChau2},
and also in the generalization to the spin-1 version of the spin-boson model \cite{HagelsteinChau3}.
In this approach, the two-level or three-level system is diagonalized directly, treating the oscillator
operators parametrically.
The rotation which accomplishes this produces weaker residual couplings when applied to the oscillator.
This rotation was discussed early on for the spin-boson problem by Wagner \cite{Wagner}.
In the multi-photon region, the energy levels of the full problem are well approximated 
away from resonance by simple averages over the parameterized diagonalized energy levels.
We expect this also to be the case in the coupled oscillator and three-level system problem.
In the first half of this paper, we develop the approximation explicitly and compare with
exact results for a test problem, with the result that the approximation gives good results
away from resonances.
We found previously in the spin-boson model that we could understand the level splittings at
resonance from a simple two basis state approximation using matrix elements of the residual
interaction arising from the rotation of the oscillator.
We are interested in whether this is also the case in the more complicated version of the problem under
consideration here.
In the second half of this paper, we develop the interaction terms explicitly and compare
approximate level splittings with exact numerical results from a test problem.
We find that the method works well for high-order resonances away from accidental or low-order
resonances on other transitions.

\newpage

\section{Basic model and unitary transformation}

The coupled oscillator and three-level system, which generalizes the spin-boson model, 
can be described using the Hamiltonian 
{\small
$$
\hat{H}
~=~
\left ( 
\begin{array} {ccc}
E_1 & 0 & 0 \cr
0 & E_2 & 0 \cr
0  & 0 & E_3 
\end{array}
\right )
+
\hbar \omega_0 \hat{a}^\dagger \hat{a}
\ \ \ \ \ \ \ \ \ \ \ \ \ \ \ \ \ \ \ \ \ \ \ \ \ \ \ 
\ \ \ \ \ \ \ \ \ \ \ \ \ \ \ \ \ \ \ \ \ \ \ \ \ \ \ 
$$
\begin{equation}
\ \ \ \ \ \ \ \ \ \ \ \ \ \ 
+
U
(\hat{a}+\hat{a}^\dagger)
\left ( 
\begin{array} {ccc}
0 & 1 & 0 \cr
1 & 0 & 0 \cr
0  & 0 & 0 
\end{array}
\right )
+
V
(\hat{a}+\hat{a}^\dagger)
\left ( 
\begin{array} {ccc}
0 & 0 & 0 \cr
0 & 0 & 1 \cr
0 & 1 & 0 
\end{array}
\right )
\end{equation}

}

\vskip 0.10in

\noindent
In this model, the three-level system has unperturbed energies $E_1$, $E_2$, and $E_3$.
Transitions between the first two levels are described by a linear coupling
term, where the interaction has strength $U$. 
Transitions between the second and third levels are also included through
another linear coupling term, with interaction strength $V$.  Such a
coupling scheme is normally called a ladder or a $\Xi$
configuration \cite{YooEberly}.  We will also assume that 

\[
|E_i - E_j| \gg \hbar \omega_0 \; , \; \; \; n \gg 1
\]

\epsfxsize = 1.80in
\epsfysize = 2.00in
\begin{figure} [t]
\begin{center}
\mbox{\epsfbox{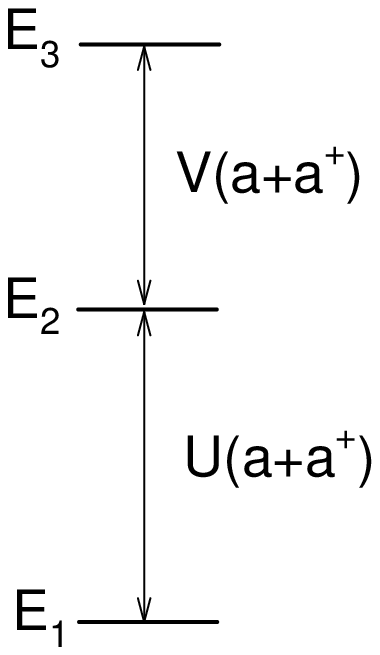}}
\caption{Three-level system and coupling.} 
\label{scheme}
\end{center}
\end{figure}

\subsection{Rotation}

In our earlier work, we used a unitary transformation to rotate the spin-boson
Hamiltonian in to a more complicated form which we wrote as

\begin{equation}
\hat{\mathcal{U}}^\dagger \hat{H} \hat{\mathcal{U}}
~=~
\hat{H}_0
~+~
\hat{V}
~+~
\hat{W}
\end{equation}

\noindent
The rotation diagonalized the two-level model parameterized by the oscillator variable $y$ defined as

\begin{equation}
y ~=~ {\hat{a} + \hat{a}^\dagger \over \sqrt{2}}
\end{equation}

\noindent
Although the transformation introduced terms more complicated mathematically, the rotated
Hamiltonian was conceptually simple. 
The first term $\hat{H}_0$ was found to produce a reasonably good approximation to the energy
levels away from the anticrossings.  
The second term $\hat{V}$ can be used to estimate the level splittings at the anticrossings.  
The third term $\hat{W}$ is small in the large $n$ regime, as is the case here.

\subsection{Unperturbed rotated Hamiltonian $\hat{H}_0$}

We focus first on $\hat{H}_0$ with the goal of developing an approximation for the
energy levels away from the anticrossings.
Following the approach used in the spin-boson version of the problem, the unitary transformation
we use here is one which diagonalizes the part of the Hamiltonian that does not include the oscillator

\begin{equation}
\hat{H}_0 - \hbar \omega_0 \hat{a}^\dagger \hat{a}
~=~
\hat{\mathcal{U}}^\dagger 
\bigg ( \hat{H} - \hbar \omega_0 \hat{a}^\dagger \hat{a} \bigg )
\hat{\mathcal{U}}
\end{equation}

\noindent
Combining the different $3 \times 3$ matrices that appear in the Hamiltonian, we may write

$$
\hat{H}_0 - \hbar \omega_0 \hat{a}^\dagger \hat{a}
~=~
\hat{\mathcal{U}}^\dagger 
\left ( 
\begin{array} {ccc}
E_1 & \sqrt{2}Uy & 0 \cr
\sqrt{2}Uy & E_2 & \sqrt{2}Vy \cr
0 & \sqrt{2}Vy  &  E_3 
\end{array}
\right )
\hat{\mathcal{U}}
\ \ \ \ \ \ \ \ \ \ \ \ 
$$
\begin{equation}
\ \ \ \ \ \ \ \ \ \ \ \ \ \ \ \ \ \ \ \ 
~=~
\left ( 
\begin{array} {ccc}
E_1(y) & 0 & 0 \cr
0 & E_2(y) & 0 \cr
0  & 0 & E_3(y) 
\end{array}
\right )
\end{equation}

\subsection{Diagonalization}

An explicit construction of the unitary transformation could be done for the simpler spin-boson problem,
but this is inconvenient for the three-level version of the problem that we are interested in here.
Nevertheless, we can still make progress by noting that the rotation accomplishes a diagonalization,
which can be done by solving

\begin{equation}
\left ( 
\begin{array} {ccc}
E_1 & \sqrt{2}Uy & 0 \cr
\sqrt{2}Uy & E_2 & \sqrt{2}Vy \cr
0 & \sqrt{2}Vy  &  E_3 
\end{array}
\right )
\left (
\begin{array} {c}
c_1 \cr
c_2 \cr
c_3 \cr
\end{array}
\right )
~=~
E(y)
\left (
\begin{array} {c}
c_1 \cr
c_2 \cr
c_3 \cr
\end{array}
\right )
\end{equation}

\noindent
This leads to a characteristic equation that is cubic

\begin{equation}
(E_1-E)(E_2-E)(E_3-E) - (E_1-E)2V^2y^2 - (E_3-E) 2 U^2 y^2 ~=~ 0
\end{equation}

\noindent
We denote the solutions to this cubic equation as $E_1(y)$, $E_2(y)$ and $E_3(y)$,
defined so that

$$
E_1(0)=E_1 
\ \ \ \ \ \ \ \ \ 
E_2(0)=E_2 
\ \ \ \ \ \ \ \ \ 
E_3(0)=E_3 
$$

\noindent
The solution of the cubic equation and expressions for the eigenvalues $E_j(y)$ are
discussed in Appendix A.
The unperturbed part of the rotated Hamiltonian $\hat{H}_0$ is then

\begin{equation}
\hat{H}_0 ~=~ 
\left ( 
\begin{array} {ccc}
E_1(y) & 0 & 0 \cr
0 & E_2(y) & 0 \cr
0  & 0 & E_3(y) 
\end{array}
\right )
~+~
\hbar \omega_0 \hat{a}^\dagger \hat{a}
\end{equation}

\subsection{Level crossing and anticrossing}

\epsfxsize = 4.50in
\epsfysize = 2.70in
\begin{figure} [t]
\begin{center}
\mbox{\epsfbox{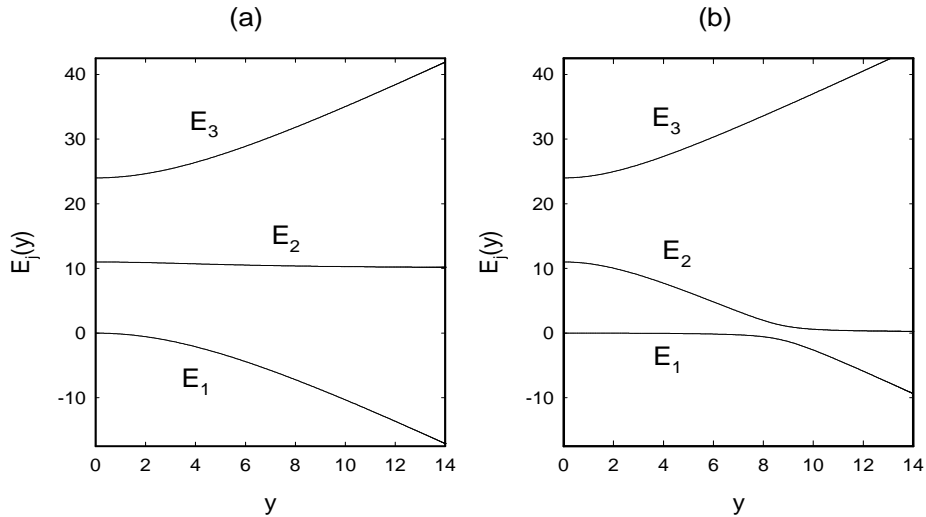}}
\caption{Energy levels $E_j(y)$ (in units of $\hbar \omega_0$) as a function of $y$ 
for a model with $E_1 = 0$, $E_2=11~ \hbar \omega_0$ and $E_3 = 24~ \hbar \omega_0$. 
The number of oscillator quanta $n$ is $10^8$.  
(a) The coupling strengths are $U\sqrt{n} = 0.8$ and $V \sqrt{n} = 0.8$.
(b) The coupling strengths are $U\sqrt{n} = 0.1$ and $V \sqrt{n} = 1.0$.
}
\label{Elevs5a}
\end{center}
\end{figure}

  The diagonalization of the three-level system parameterized by $y$ produces energy levels that
depend on $y$, examples of which are illustrated in Figure \ref{Elevs5a}.
We expect static coupling between levels to push them apart, and in the calculation of Figure \ref{Elevs5a}(a)
this is exactly what we see.
However, since the interaction pushes levels apart, it is possible that one level may be pushed into and
through another level.
This is illustrated in Figure \ref{Elevs5a}(b), where the interaction between levels 2 and 3 is strong,
and level one is weakly coupled.
In this case level 2 anticrosses level 1.
We refer to the middle level as $E_2(y)$ on either side of the anticrossing.

\newpage
\section{Energy levels away from anticrossings}

\epsfxsize = 3.60in
\epsfysize = 2.80in
\begin{figure} [t]
\begin{center}
\mbox{\epsfbox{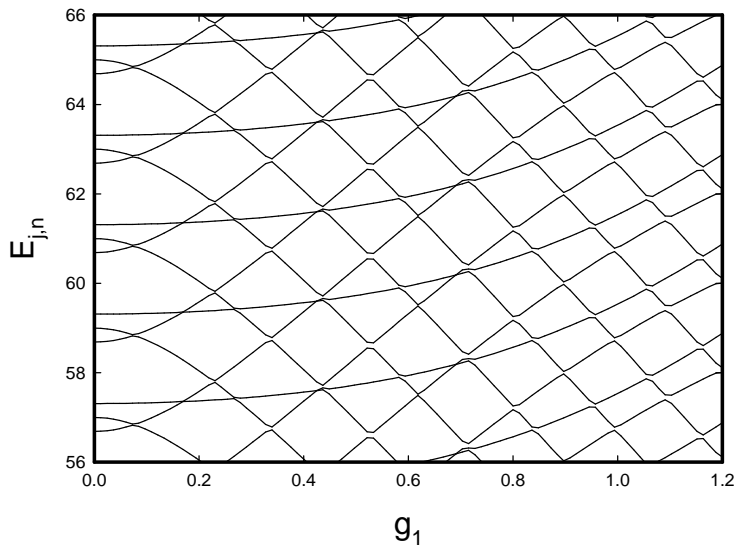}}
\caption{Energy levels as a function of $g_1$ for $g_2=1/3$ for even parity states, in the case of a model
with $E_1 = 0$, $E_2 = 11 ~ \hbar \omega_0$, and $E_3 = 24 ~ \hbar \omega_0$, with $n_0 = 10^8$.
The energy levels are in units of $\hbar \omega_0$, and are offset by $10^8~ \hbar \omega_0$.} 
\label{ws}
\end{center}
\end{figure}

To understand the systematics of the energy levels we consider solutions to the time-independent
Schr\"odinger equation

\begin{equation}
E \Psi ~=~ \hat{H} \Psi
\end{equation}

\noindent
Some of the energy levels from such a computation are illustrated in Figure \ref{ws}, where we have 
defined the dimensionless coupling strengths $g_1$ and $g_2$ according to

\begin{equation}
g_1 ~=~ {U \sqrt{n} \over E_2-E_1}
\ \ \ \ \ \ \ \ \
g_2 ~=~ {V \sqrt{n} \over E_3 - E_2}
\end{equation}

\noindent
There are many energy levels; we expect three levels for each $n$, which
is the same as 
we would get in the absence of coupling.  As one can see from Figure
\ref{ws},  the complete spectrum is locally made up of three
curves with  
the same pattern repeated over and over again with increasing 
energy (where the energy offset is $2 ~ \hbar \omega_0$).  The actual
offset is $\hbar \omega_0$, but we have only included the even parity
(even total index $j$ plus $n$)  energy levels to make the figure simpler.
The energy levels of the odd parity (odd total index $j$ plus $n$) are
nearly identical, shifted by one unit of $\hbar \omega_0$.  This
periodicity occurs because in the large $n$ limit, the coupling of $n + 1$
quanta is little different from coupling of $n$ quanta.  Under these
conditions, the energy levels can be parameterized according to 

\begin{equation}
E_{j,n}(g_1,g_2) ~=~ E_j(g_1,g_2) + n \hbar \omega_0
\end{equation}

\noindent
The corresponding physical statement is that the oscillator impacts the three-level system
strongly, causing the levels to shift; while the three-level system has very little impact
on average on the oscillator, so that its levels are not shifted.
This situation is similar to that observed in the spin-boson problem when $n$ is large \cite{HagelsteinChau2}.
This motivates us to try to understand the shifted (or dressed) energy levels $E_j(g_1,g_2)$,
as these are fundamental to the model in this limit.

\subsection{Dressed energy levels from the rotated Hamiltonian $\hat{H}_0$}

We can develop an effective estimate for the energy levels away from
the anticrossings using an approach similar to that used for the spin-boson
problem.
We base our approximation on the eigenvalues of the rotated $\hat{H}_0$ problem
which can be written as

{\small

\begin{equation}
\left (E + {1 \over 2} \hbar \omega_0 \right ) \Phi
~=~
\left ( 
\begin{array} {ccc}
E_1(y) & 0 & 0 \cr
0 & E_2(y) & 0 \cr
0  & 0 & E_3(y) 
\end{array}
\right )
+
{1 \over 2} \hbar \omega_0
\left ( 
-{d^2 \over dy^2} + y^2 
\right )
\Phi
\end{equation}

}

\noindent
We can separate the three-level degrees of freedom from the oscillator degree of freedom
by using wavefunctions of the form

{\small

\begin{equation}
\Phi_{1,n}
= 
\left (
\begin{array} {c}
1 \cr
0 \cr
0 \cr
\end{array}
\right )
u_1(y)
\ \ \ \ \ \
\Phi_{2,n}
= 
\left (
\begin{array} {c}
0 \cr
1 \cr
0 \cr
\end{array}
\right )
u_2(y)
\ \ \ \ \ \
\Phi_{3,n}
= 
\left (
\begin{array} {c}
0 \cr
0 \cr
1 \cr
\end{array}
\right )
u_3(y)
\end{equation}

}

\noindent
The $u_j(y)$ functions satisfy a one-dimensional Schr\"odinger equation of the form

\begin{equation}
\left (E + {1 \over 2} \hbar \omega_0 \right ) u_j(y)
~=~
E_j(y)
+
{1 \over 2} \hbar \omega_0
\left (
-{d^2 \over dy^2} + y^2 
\right )
\label{H0problem}
\end{equation}

\noindent
This is similar to what we obtained in the spin-boson
problem \cite{HagelsteinChau2}.

\subsection{Variational estimate}

A useful approximation valued away from the level anticrossings can be developed by
adopting trial wavefunctions in which the $u_j(y)$ function is taken to be a simple
harmonic oscillator functions

\begin{equation}
u_j(y) ~=~ \phi_n(y) 
\end{equation}

\noindent
Such a wavefunction is relevant in the large $n$ limit for a variational estimate of
the energy, leading to the approximation

\begin{equation}
E_{j,n}(g_1,g_2) 
~=~ 
\langle n |E_j(y) |n \rangle + n \hbar \omega_0
~=~
E_j(g_1,g_2) + n \hbar \omega_0
\label{variational}
\end{equation}

\noindent
Away from the anticrossings, the eigenfunctions of the original Hamiltonian $\hat{H}$
have average $n$ and $j$ values which are close to the integer values which we assign
in the rotated $\hat{H}_0$ problem here.  
When $n$ is large, we can take advantage of the WKB approximation to write

\begin{equation}
\langle n |E_j(y) |n \rangle
~=~
{1 \over \pi} \int_{-\epsilon}^\epsilon {E_j(y) \over \sqrt{\epsilon - y^2}} dy
\label{WKB}
\end{equation}

\noindent
with $\epsilon = 2n+1$.  
When $n$ is large, the WKB approximation gives excellent agreement with the results
from a brute force solution of Equation (\ref{H0problem}), as was the case previously
for the spin-boson problem.
An algebraic expression in the case of the middle level is given in the Appendix.

\newpage
\section{Model test problem}

To illustrate the approximation under discussion, we consider a specific example.
In this example, we are interested in the basic question of whether the approximation
described in the previous section is a good one away from resonance.
However, since the three-level model and coupled oscillator problem is considerably 
more complicated than the analogous spin-boson problem, our task becomes more
difficult.
We focus on a three-level system with unperturbed energy splittings given by

\begin{equation}
E_2-E_1 = 11 ~\hbar \omega_0
\ \ \ \ \ \ \ \ \ \
E_3-E_2 = 13 ~\hbar \omega_0
\end{equation}

\noindent
with $g_1$ and ranging between 0 and 1, and $g_2$ ranging between 0 and 1.2 (we 
extend the range for $g_2$ since interesting features extend out a bit further
along the $g_2$ axis).  

\subsection{Resonances associated with transitions between levels 1 and 2}

We expect that the approximation for the energy levels discussed above will be
best away from resonances, so our first goal is to understand where resonances
occur. 
The most important resonances are those involving transitions between levels 1 and 2,
and levels 2 and 3.
In the case of resonances involving the lowest two levels, the resonance conditions 
is given approximately by

\begin{equation}
E_2(g_1,g_2) - E_1(g_1,g_2) ~=~ \Delta n \hbar \omega_0
\end{equation}

\noindent
where $\Delta n$ is odd.
The associated contours are illustrated as a function of the dimensionless coupling
constants $g_1$ and $g_2$ in Figure \ref{ee21}, based on the WKB approximation.
When both $g_1$ and $g_2$ are small, it should take 11 oscillator quanta to match
the transition energy since the energy levels in this case are only weakly perturbed.
For small $g_2$, the dressed transition energy $E_2(g_1,g_2) - E_1(g_1,g_2)$ increases with increasing $g_1$,
so that more oscillator quanta are required for a resonance.
This can be seen in Figure \ref{ee21} in the resonance contours for increasing $\Delta n$
coming down to the $g_1$ axis.
In the upper left of this figure, things are much more complicated.
As $g_2$ increases, the splitting between the upper two levels is increased.
If $g_1$ is small, then level 1 is only weakly coupled, and it is possible for 
level 2 to be pushed near level 1.
The separation between the lower two levels is decreased, so that lower-order
resonances (in which $\Delta n$ is smaller) occur.
The coupling due to the $\hat{V}$ operator between the two lowest levels 
is much stronger for these lower-order resonances, so that the mixing is
much greater and the associated level splitting is much larger.
Approximating these states as pure eigenfunctions of the $\hat{H}_0$
Hamiltonian becomes a much poorer approximation in this regime.

\begin{samepage}

\epsfxsize = 4.08in
\epsfysize = 2.90in
\begin{figure} [t]
\begin{center}
\mbox{\epsfbox{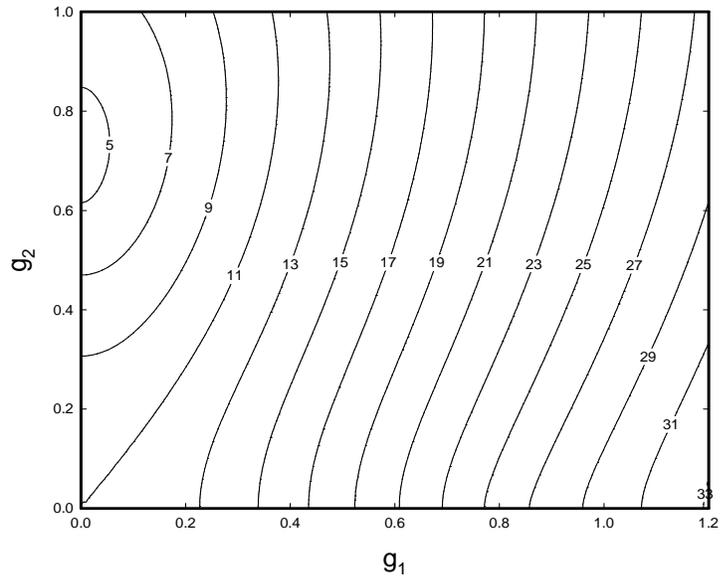}}
\caption{
Contours for resonances between levels 2 and 1 (from the rotated $\hat{H}_0$ Hamiltonian), 
with the exchange of an odd number of oscillator quanta. The number of oscillator quanta $\Delta n$
is indicated with each resonance.} 
\label{ee21}
\end{center}
\end{figure}

\epsfxsize = 4.80in
\epsfysize = 2.90in
\begin{figure} [t]
\begin{center}
\mbox{\epsfbox{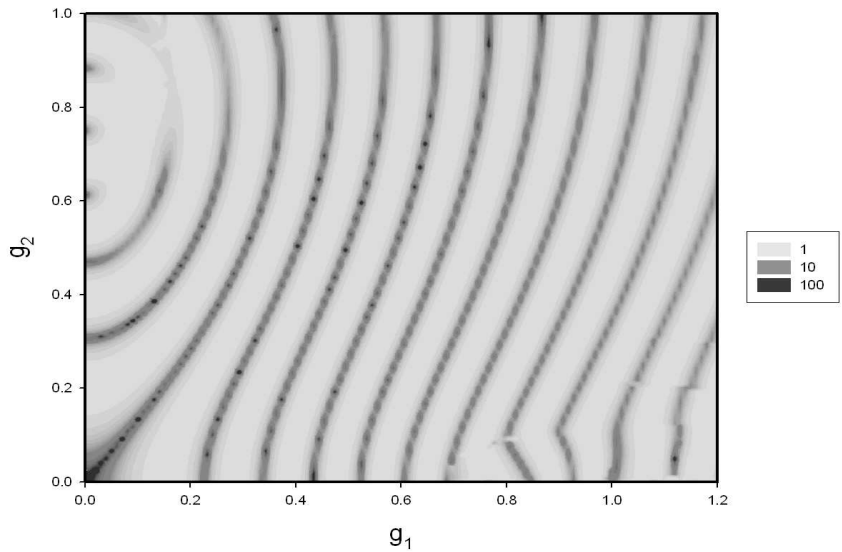}}
\caption{Resonances on transitions between levels 1 and 2 through the inverse magnitude of
the resonance condition $|E_2(g_1,g_2)-E_1(g_1,g_2)-\Delta n \hbar \omega_0|^{-1}$  
as a function of $g_1$ and $g_2$. 
Computations for this plot are done using the original unrotated Hamiltonian $\hat{H}$.
The results are in units of $(\hbar \omega_0)^{-1}$.} 
\label{eres21}
\end{center}
\end{figure}

\end{samepage}

We can see the resonances in the results shown in Figure \ref{eres21} taken
from a direct computation using the original $\hat{H}$ Hamiltonian.
One can see that most of the resonances predicted by the WKB approximation
appear in the full computation.

\subsection{Resonances associated with transitions between levels 2 and 3}

\begin{figure} [t]
\begin{center}
\mbox{\epsfbox{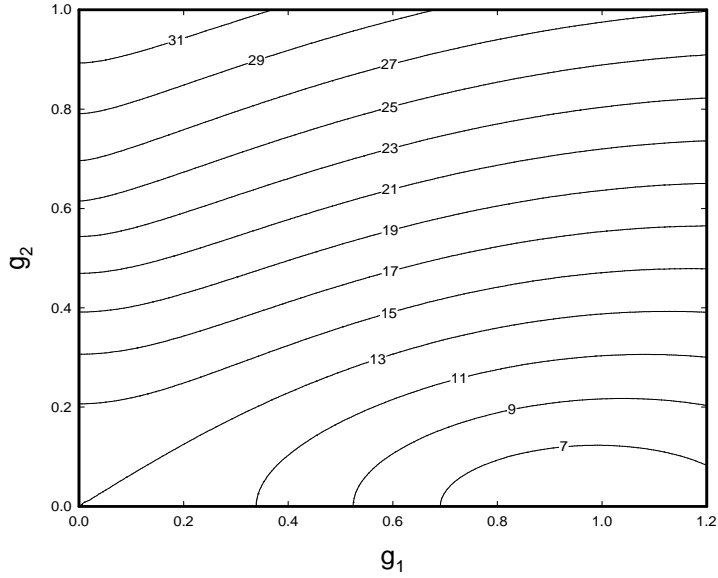}}
\caption{
Contours for resonances between levels 3 and 2 (from the rotated $\hat{H}_0$ Hamiltonian), 
with the exchange of an odd number $\Delta n$ of oscillator quanta.  The contour labels indicate $\Delta n$.} 
\label{ee32}
\end{center}
\end{figure}

The resonance condition associated with transitions between levels 2 and 3 is
given approximately by

\begin{equation}
E_3(g_1,g_2) - E_2(g_1,g_2) ~=~ \Delta n \hbar \omega_0
\end{equation}

\noindent
where once again the number of oscillator quanta exchanged must be odd.
This resonance condition from WKB calculations is illustrated in Figure \ref{ee32}.
For small $g_1$ and $g_2$ it now takes 13 oscillator quanta to match the
transition energy, since in this case it is only weakly perturbed.
When $g_1$ is small, an increase in $g_2$ pushes levels 2 and 3 apart,
increasing the number of quanta required for a resonance.
This situation is qualitatively similar as considered above, except
that the axes are reversed.
When $g_2$ is small, large values of $g_1$ can cause levels 1 and 2 to
push apart strongly, driving level 2 toward level 3.
Due to the asymmetry between the level splittings, this occurs 
for $g_1$ values larger than those for $g_2$ in the previous case
for resonances between levels 1 and 2.
The resulting small separation that occurs in the lower right hand part
of Figure \ref{ee32} between levels 2 and 3 leads to the occurrence of
low-order resonances, with the associated strong mixing and large
level shifts due to the $\hat{V}$ operator.

\begin{figure} [t]
\begin{center}
\mbox{\epsfbox{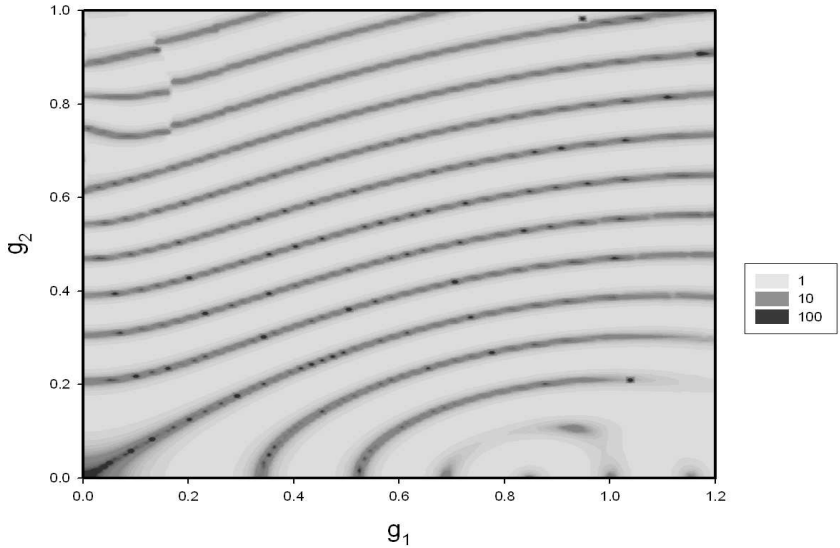}}
\caption{Resonances on transitions between levels 1 and 2 through the inverse magnitude of
the resonance condition $|E_3(g_1,g_2)-E_2(g_1,g_2)-\Delta n \hbar \omega_0|^{-1}$  
as a function of $g_1$ and $g_2$. 
Computations for this plot are done using the original unrotated Hamiltonian $\hat{H}$.
The results are in units of $(\hbar \omega_0)^{-1}$.} 
\label{eres32}
\end{center}
\end{figure}

Once again we show analogous results from direct calculations based on
the original $\hat{H}$ Hamiltonian in Figure \ref{eres32}.
Most of the resonance lines from the WKB calculation are evident in
the full $\hat{H}$ calculation.
The effects of the low-order resonances between levels 3 and 2 are
apparent in the lower right part of the plot.

\subsection{Energy levels}

\epsfxsize = 4.08in
\epsfysize = 2.90in

\epsfxsize = 4.80in
\epsfysize = 2.90in

\begin{figure} [t]
\begin{center}
\mbox{\epsfbox{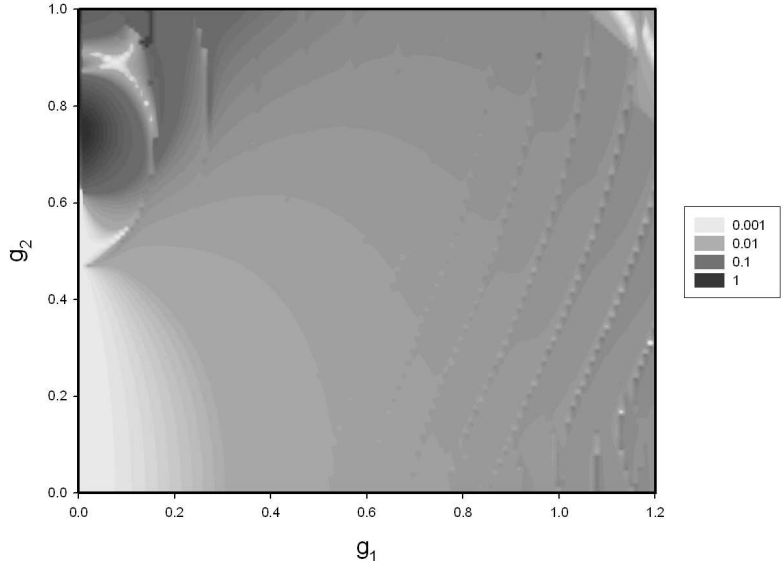}}
\caption{Magnitude of difference between the exact result for $E_1(g_1,g_2)$
from numerical calculations of the unperturbed Hamiltonian $\hat{H}$, and
the WKB approximation for $E_1(g_1,g_2)$ in units of $\hbar \omega_0$.}
\label{de1}
\end{center}
\end{figure}

From the discussion above, we expect that the WKB energy level estimates
based on the rotated Hamiltonian $\hat{H}_0$ should be good away from regions
where resonances occur, and should be best where the separation between
the nearest level is largest.
In Figure \ref{de1} we illustrate the magnitude of the difference between
the exact energies level 1 obtained from the original $\hat{H}$ problem, 
and the WKB approximation:

$$
\left |
E_1(g_1,g_2) - {1 \over \pi} \int_{-\epsilon}^\epsilon {E_1(y) \over \sqrt{\epsilon - y^2}} dy
\right |
$$

\noindent
One sees that when $g_1$ and $g_2$ are small, that the difference is also small,
and the approximation is very good.
In the upper left part of the plot there occur the largest deviations as we expected
due to low-order mixing effects not included in the $\hat{H}_0$ problem.
Elsewhere the agreement is generally rather good away from the resonances.
The energy shift associated with the high-order resonances is not so large,
and deviations are observable only when $g_1$ becomes large.
We note that the energy level separation between the lowest two levels
away from the upper left part of the plot is between about 10 and 40
$\hbar \omega_0$; hence the deviation between exact and approximate
results is generally less than 0.1\% of this separation.
In the region where we would expect the approximation to be relevant, we
find that it is very good.

\epsfxsize = 4.38in
\epsfysize = 3.00in

\begin{figure} [t]
\begin{center}
\mbox{\epsfbox{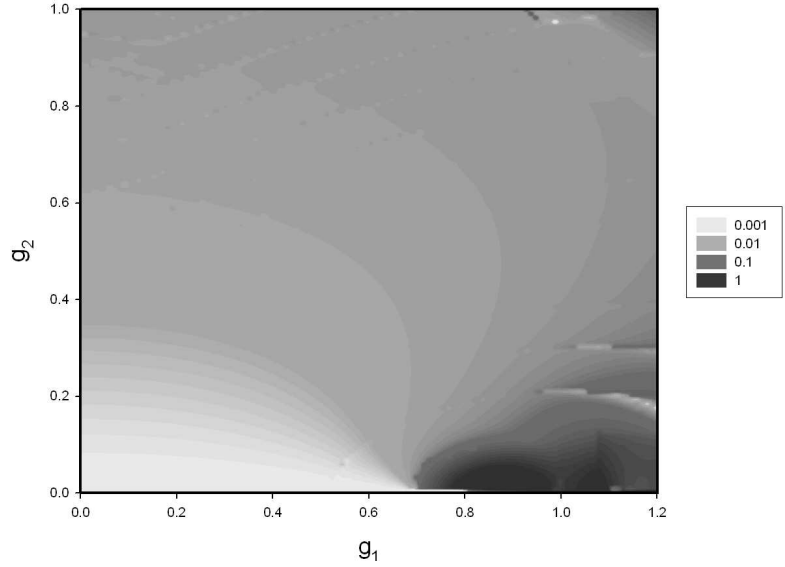}}
\caption{Magnitude of difference between the exact result for $E_3(g_1,g_2)$
from numerical calculations of the unperturbed Hamiltonian $\hat{H}$, and
the WKB approximation for $E_3(g_1,g_2)$ in units of $\hbar \omega_0$.}
\label{de3}
\end{center}
\end{figure}

In Figure \ref{de3} we show similar results for the magnitude of the difference
in energy for level 3.
Once again, the difference is small when $g_1$ and $g_2$ are small.
Away from the lower right region (where level 2 is pushed toward level 3)
the differences are small and the approximation is good.
In the lower right region, strong mixing occurs due to the low-order resonances,
and the approximation under discussion degrades.

\begin{figure} [t]
\begin{center}
\mbox{\epsfbox{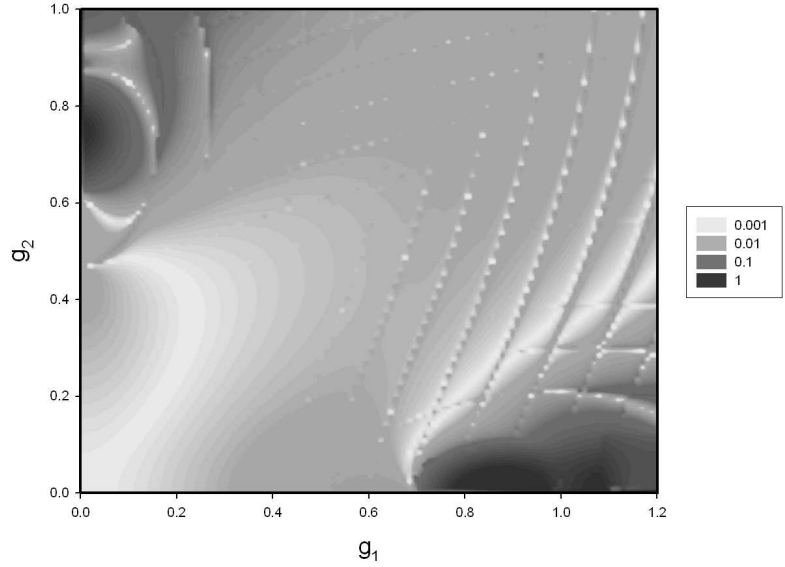}}
\caption{Magnitude of difference between the exact result for $E_2(g_1,g_2)$
from numerical calculations of the unperturbed Hamiltonian $\hat{H}$, and
the WKB approximation for $E_2(g_1,g_2)$ in units of $\hbar \omega_0$.}
\label{de2}
\end{center}
\end{figure}

In the case of level 2, the region where the approximation is accurate becomes further
restricted, as can be seen in Figure \ref{de2}.
The strong mixing that we previously encountered in the upper left part of the plot
in the case of level 1 is reproduced in the case of level 2, since level 1 is mixing
primarily with level 2.
The strong mixing associated with level 3 in the lower right part of the plot in the
case of level 3 is also reproduced in the case of level 2, since level 3 is mixing
with level 2 as well.
Consequently, the best results are obtained in the central region, away from the
problem regions in the upper left and lower right part of the region.

\epsfxsize = 4.38in
\epsfysize = 3.00in

\epsfxsize = 4.38in
\epsfysize = 3.00in

\newpage

\section{Remaining terms of the rotated Hamiltonian}

We set out initially to implement a unitary transformation similar to the
one which we studied previously in the spin-boson problem, in order to
obtain a rotated Hamiltonian of the form

$$
\hat{\mathcal{U}}^\dagger \hat{H} \hat{\mathcal{U}}
~=~
\hat{H}_0
~+~
\hat{V}
~+~
\hat{W}
$$

\noindent
In previous sections we developed the first term in the rotated Hamiltonian $\hat{H}_0$,
and compared results from computations with those derived from the original Hamiltonian $\hat{H}$.
To complete the unitary transformation, we need to rotate the simple harmonic
oscillator part of the Hamiltonian to obtain $\hat{V}$ and $\hat{W}$.
We may write

\begin{equation}
\hat{V} + \hat{W}  
~=~ 
\hat{\mathcal{U}}^\dagger \bigg [ \hbar \omega_0 \hat{a}^\dagger \hat{a} \bigg ] \hat{\mathcal{U}}
-
\hbar \omega_0 \hat{a}^\dagger \hat{a}
\end{equation}

\subsection{Transformation under rotation}

Perhaps the simplest way to approach this problem is to examine how the simple harmonic
oscillator operators transform under the unitary transform.
The oscillator operators under discussion here are $y$, and $\displaystyle{d \over dy}$, 
where

$${d \over dy} ~=~ {\hat{a} - \hat{a}^\dagger \over \sqrt{2}}$$

\noindent
Since the rotation diagonalizes the three-level system for each $y$, the unitary operators
itself depends on $y$.
We may write

$$\hat{\mathcal{U}} ~=~ \hat{\mathcal{U}}(y)$$

\noindent
to exhibit this explicitly. 
There is no dependence of the unitary operator on $\displaystyle{d \over dy}$.
Consequently, the rotation of the oscillator operators can be written as

\begin{equation} 
y' ~=~ \hat{\mathcal{U}}^\dagger(y) y \hat{\mathcal{U}}(y) ~=~ y
\end{equation}

\begin{equation} 
{d \over dy'} 
~=~ 
\hat{\mathcal{U}}^\dagger(y) {d \over dy} \hat{\mathcal{U}}(y) 
~=~ {d \over dy} + \hat{\mathcal{U}}^\dagger(y) \left [ {d \over dy} \hat{\mathcal{U}}(y) \right ] 
\end{equation}

\subsection{Rotation of the harmonic oscillator}

  We can make use of this transformation to construct the other terms in the rotated Hamiltonian.
We have

\begin{equation}
\hat{\mathcal{U}}^\dagger \bigg [ \hat{a}^\dagger \hat{a} \bigg ] \hat{\mathcal{U}}
~=~
\hat{\mathcal{U}}^\dagger 
\left [ {1 \over 2} \bigg ( y^2 - {d^2 \over dy^2} - 1\bigg )  \right ] 
\hat{\mathcal{U}}
~=~
{1 \over 2} \bigg ( (y')^2 - \left [{d \over dy'}\right ]^2 - 1 \bigg )
\end{equation}

\noindent
Inserting expressions for the transformed operators produces


{\small

$$
\hat{\mathcal{U}}^\dagger \bigg [ \hat{a}^\dagger \hat{a} \bigg ] \hat{\mathcal{U}}
-
\hat{a}^\dagger \hat{a}
~=~
\ \ \ \ \ \ \ \ \ \ \ \ \ \ \ \ \ \ \ \ \ \ \ \ \ \ \ \ \ \ \ \ \ \ \ \ \ \ \ \ \ \
$$
\begin{equation}
\ \ \ \ \ \ \
-
{1 \over 2} \left \lbrace
{d \over dy} \left [ \hat{\mathcal{U}}^\dagger(y){d \over dy} \hat{\mathcal{U}}(y) \right ]
+
\left [ \hat{\mathcal{U}}^\dagger(y){d \over dy} \hat{\mathcal{U}}(y) \right ]{d \over dy}
+
\left [ \hat{\mathcal{U}}^\dagger(y){d \over dy} \hat{\mathcal{U}}(y) \right ]^2
\right \rbrace
\end{equation}

}

\noindent
This allows us to identify the two remaining parts of the rotated Hamiltonian, which we may
write as

\begin{equation}
\hat{V} 
~=~ 
-
{\hbar \omega_0 \over 2} \left \lbrace
{d \over dy} \left [ \hat{\mathcal{U}}^\dagger(y){d \over dy} \hat{\mathcal{U}}(y) \right ]
+
\left [ \hat{\mathcal{U}}^\dagger(y){d \over dy} \hat{\mathcal{U}}(y) \right ]{d \over dy}
\right \rbrace
\end{equation}

\begin{equation}
\hat{W} ~=~
-
{\hbar \omega_0 \over 2} 
\left [ \hat{\mathcal{U}}^\dagger(y){d \over dy} \hat{\mathcal{U}}(y) \right ]^2
\end{equation}

\subsection{Reduction of $\hat{\mathcal{U}}^\dagger(y)\displaystyle{d \over dy} \hat{\mathcal{U}}(y)$}

It is possible to simplify these operators by taking advantage of properties of the
unitary matrix.
The unitary operator satisfies

\begin{equation}
\hat{\mathcal{U}}^\dagger(y)\hat{\mathcal{U}}(y) ~=~ 1
\end{equation}

\noindent
independent of the parameter $y$.  Hence, if we differentiate in $y$ we obtain

\begin{equation}
{d \over dy} [\hat{\mathcal{U}}^\dagger(y)\hat{\mathcal{U}}(y)
~=~
\left [{d \over dy} \hat{\mathcal{U}}^\dagger(y) \right ]\hat{\mathcal{U}}(y)
+
\hat{\mathcal{U}}^\dagger(y) \left [{d \over dy} \hat{\mathcal{U}}(y) \right ]
~=~ 0
\end{equation}

\noindent
A consequence of this is that the matrix which represents 
$\hat{\mathcal{U}}^\dagger(y)\displaystyle{d \over dy} \hat{\mathcal{U}}(y)$
is antisymmetric if we chose $\mathcal{U}(y)$ to be a real matrix.
In this case we may write

\begin{equation}
\left (
\begin{array} {ccc}
\mathcal{U}_{11} & \mathcal{U}_{21} & \mathcal{U}_{31} \cr
\mathcal{U}_{12} & \mathcal{U}_{22} & \mathcal{U}_{32} \cr
\mathcal{U}_{13} & \mathcal{U}_{23} & \mathcal{U}_{33} \cr
\end{array}
\right )
\left (
\begin{array} {ccc}
\mathcal{U}_{11}' & \mathcal{U}_{12}' & \mathcal{U}_{13}' \cr
\mathcal{U}_{21}' & \mathcal{U}_{22}' & \mathcal{U}_{23}' \cr
\mathcal{U}_{31}' & \mathcal{U}_{32}' & \mathcal{U}_{33}' \cr
\end{array}
\right )
~=~
\left (
\begin{array} {ccc}
0       & F_{12} & F_{13} \cr
-F_{12} & 0      & F_{23} \cr
-F_{13} & F_{23} & 0      \cr
\end{array}
\right )
\end{equation}

\noindent
where

$$\mathcal{U}_{ij}' ~=~ {d \over dy} \mathcal{U}_{ij}(y)$$

\noindent
The $F_{ij}$ matrix elements can be obtained directly from the unitary matrix,
which allows us to write

\begin{samepage}

\begin{equation}
F_{12} ~=~ 
\mathcal{U}_{11} \mathcal{U}_{12}' 
+
\mathcal{U}_{21} \mathcal{U}_{22}' 
+
\mathcal{U}_{31} \mathcal{U}_{32}'
\end{equation}

\begin{equation}
F_{23} ~=~ 
\mathcal{U}_{12} \mathcal{U}_{13}' 
+
\mathcal{U}_{22} \mathcal{U}_{23}' 
+
\mathcal{U}_{32} \mathcal{U}_{33}'
\end{equation}

\begin{equation}
F_{13} ~=~ 
\mathcal{U}_{11} \mathcal{U}_{13}' 
+
\mathcal{U}_{21} \mathcal{U}_{23}' 
+
\mathcal{U}_{31} \mathcal{U}_{33}'
\end{equation}

\end{samepage}

\noindent
The $\hat{\mathcal{U}}^\dagger(y)\displaystyle{d \over dy} \hat{\mathcal{U}}(y)$
operator can then be expanded as a sum of spatial parts and SU(3) matrices according to

{\small

$$
\hat{\mathcal{U}}^\dagger(y){d \over dy} \hat{\mathcal{U}}(y)
~=~
i F_{12}(y)
\left (
\begin{array} {ccc}
0 & -i & 0 \cr
i &  0 & 0 \cr
0 &  0 & 0 \cr
\end{array}
\right )
+
i F_{13}(y)
\left (
\begin{array} {ccc}
0 & 0 & -i \cr
0 &  0 & 0 \cr
i &  0 & 0 \cr
\end{array}
\right )
$$
\begin{equation}
\ \ \ \ \ \ \ \ \ \ \ \ \ \ \ \ \ \ \ \
+
i F_{23}(y)
\left (
\begin{array} {ccc}
0 &  0 &  0 \cr
0 &  0 & -i \cr
0 &  i &  0 \cr
\end{array}
\right )
\end{equation}

}

\subsection{Reduction of $\hat{V}$}

We can use these results to recast the $\hat{V}$ in a similar form.
We may write

{\small

$$
\hat{V}
~=~
-i
{\hbar \omega_0 \over 2} 
\left \lbrace
\left [
{d \over dy} 
F_{12}(y)
+
F_{12}(y)
{d \over dy}
\right ]
\left (
\begin{array} {ccc}
0 & -i & 0 \cr
i &  0 & 0 \cr
0 &  0 & 0 \cr
\end{array}
\right )
\right .
\ \ \ \ \ \ \ \ \ \ \ \ \ \ \ \ \ \ \ \ \ \ \ \ \ \
$$
$$
+
\left [
{d \over dy} 
F_{13}(y)
+
F_{13}(y)
{d \over dy}
\right ]
\left (
\begin{array} {ccc}
0 & 0 & -i \cr
0 &  0 & 0 \cr
i &  0 & 0 \cr
\end{array}
\right )
$$
\begin{equation}
\ \ \ \ \ \ \ \ \ \ \ \ \ \ \ \ \ \ \ \ \ \ \ \ \ \ \
\left .
+
\left [
{d \over dy} 
F_{23}(y)
+
F_{23}(y)
{d \over dy}
\right ]
\left (
\begin{array} {ccc}
0 &  0 &  0 \cr
0 &  0 & -i \cr
0 &  i &  0 \cr
\end{array}
\right )
\right \rbrace
\end{equation}

}

\noindent
The $\hat{V}$ operator written in this form is seen to be closely related
to the one obtained in the spin-boson problem.
A similar reduction of $\hat{W}$ can be carried out without difficulty from
these results.

\newpage

\section{Level splitting at the anticrossings}

As discussed above, the energy levels away from the anticrossings are given approximately by

$$
E_{j,n} ~=~ E_j(g_1,g_2) + n \hbar \omega_0
$$

\noindent
The approximate resonance conditions for the level anticrossings between the lowest two
level as discussed above is

$$
E_{2}(g_1,g_2)-E_1(g_1,g_2) ~=~ \Delta n \, \hbar \omega_0
$$

\noindent
for $\Delta n$ odd.
When we studied the rotated spin-boson system using a similar unitary transformation,
we found that the level splitting was well approximated using degenerate perturbation
theory in the rotated frame based on matrix elements of the $\hat{V}$ operator.
We expect a similar behavior in the three-level version of the problem, which would
result in an estimate for the level splitting given by

\begin{equation}
\delta E ~=~ 2 | \langle \Phi_{1,n} | \hat{V} | \Phi_{2, n - \Delta n} \rangle |
\end{equation}

\subsection{Level splittings, large $\Delta n$}

As discussed in Section 4, the regions with different $g_1$ and $g_2$ would be expected
to behave differently due to the presence of low-order resonances.
Hence, we begin the discussion with a comparison in a relatively benign region in which
levels 1 and 2 have large separation, and we stay away from the problem area in the
lower right part of the plots discussed in Section 4.
We seek resonances along a line with

$$g_2 ~=~ 0.3 ~g_1$$

\begin{figure} [t]
\begin{center}
\mbox{\epsfbox{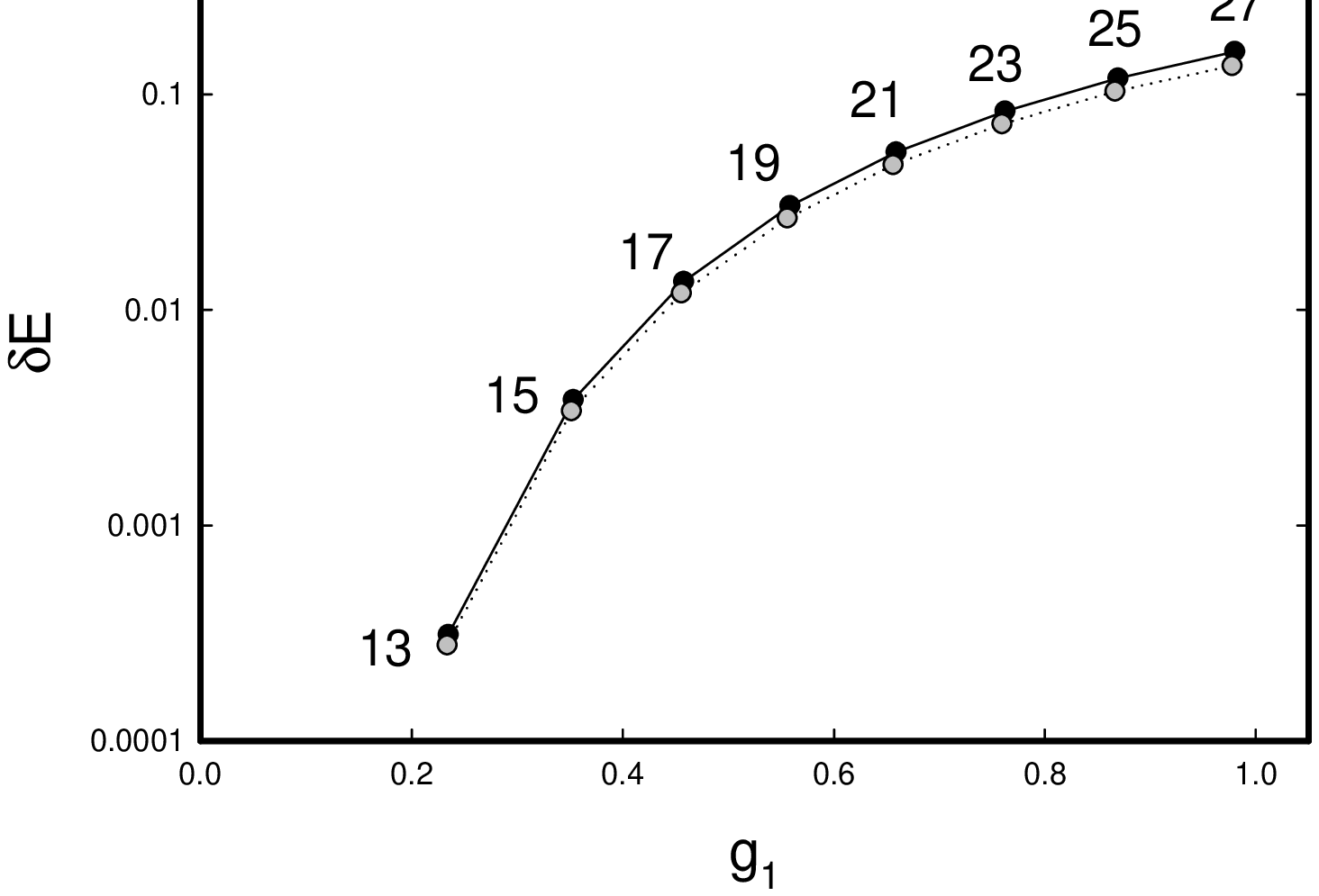}}
\caption{Level splitting at the anticrossings between levels 1 and 2 for the model
described in Section 4, with $g_2 = 0.3 g_1$.  Black circles: degenerate
perturbation theory based on wavefunctions in the rotated frame; gray circles: results from a
direct computation using the original unrotated Hamiltonian $\hat{H}$. Energies are given in units
of $\hbar \omega_0$. }
\label{spl12x03}
\end{center}
\end{figure}

\noindent
Along this line the resonances between the lower two levels do not overlap other resonances;
the resonances are all of high order, and there is no interference from the problem area
associated with levels 2 and 3.
Consequently we would expect the lower two levels to give good agreement, much as we found
in the spin-boson problem with a two-level system.
Results from computations based on the model test problem discussed in Section 4 are
shown in Figure \ref{spl12x03}.
We see in this figure very good agreement, similar to what we found in the spin-boson case \cite{HagelsteinChau2}.

\epsfxsize = 3.60in
\epsfysize = 2.80in

\subsection{Interaction with low-order resonances}

\begin{figure} [t]
\begin{center}
\mbox{\epsfbox{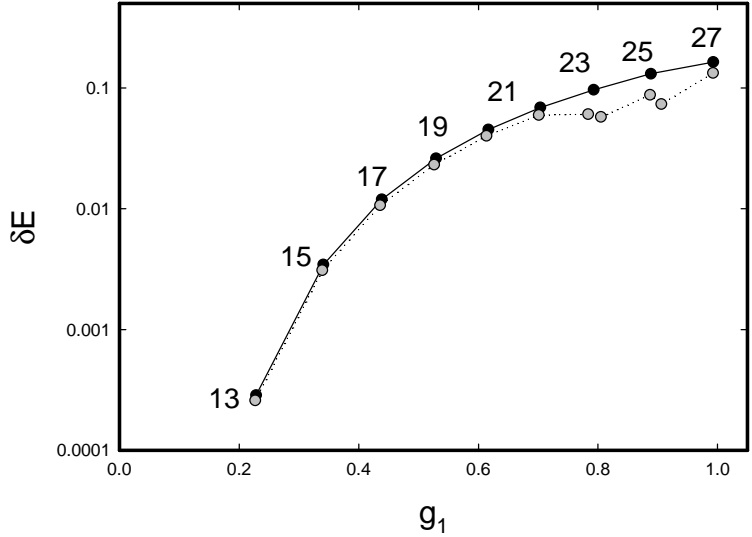}}
\caption{Level splitting at the anticrossings between levels 1 and 2 for the model
described in Section 4, with $g_2 = 0.1 g_1$.  Black circles: degenerate
perturbation theory based on wavefunctions in the rotated frame; gray circles: results from a
direct computation using the original unrotated Hamiltonian $\hat{H}$. Energies are given in units
of $\hbar \omega_0$. }
\label{spl12x01}
\end{center}
\end{figure}

We would expect a degradation of the approximation in the vicinity of the region where
the other transition experiences low-order resonances.
We continue to focus on the resonances involving the lowest two levels, this time on a line
given by

$$g_2 ~=~ 0.1 ~g_1$$

\noindent
which goes into the problem area with low-order transitions between the upper two 
levels.
Results are presented in Figure \ref{spl12x01}.
We see in this case good agreement between the results from degenerate perturbation theory based on
the rotated problem, and results from the original $\hat{H}$ problem everywhere except at resonances
with $\Delta n$ equal to 23 and 25.
At these points the upper two levels are mixed due to strong low-order $\hat{V}$ interactions, and
one finds two resonances in the vicinity of the single resonance predicted by the rotated $\hat{H}_0$
model.

\epsfxsize = 3.60in
\epsfysize = 2.80in

\subsection{Other effects}

\begin{figure} [t]
\begin{center}
\mbox{\epsfbox{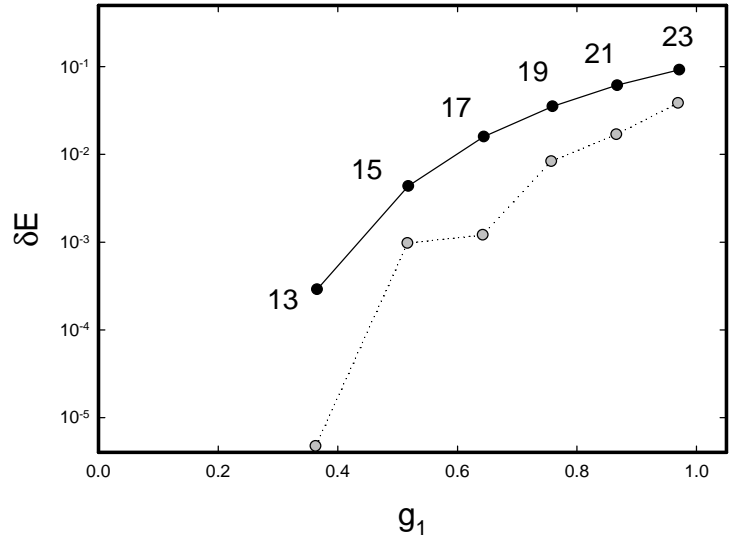}}
\caption{Level splitting at the anticrossings between levels 1 and 2 for the model
described in Section 4, with $g_2 = 1.1 g_1$.  Black circles: degenerate
perturbation theory based on wavefunctions in the rotated frame; gray circles: results from a
direct computation using the original unrotated Hamiltonian $\hat{H}$. Energies are given in units
of $\hbar \omega_0$. }
\label{spl12x11}
\end{center}
\end{figure}

We find poorer agreement when we move into the region in which $g_2$ is greater than $g_1$.
We consider in Figure \ref{spl12x11} resonances along the line defined by

$$g_2 = 1.1 ~g_1$$

\noindent
The results in this case can be understood as being due to two different effects.
The level splittings at the anticrossings for the full $\hat{H}$ problem begin to fall 
systematically below those obtained from degenerate perturbation theory for the rotated 
problem once we consider the $g_2=g_1$ line.
Two of the resonances (13 and 17) are lower still due to interference from nearby
resonances involving the upper two levels.
We find poorer agreement between the exact numerical results for the $\hat{H}$
problem and those from the approximation in this region, 
and worse still as the ratio of $g_2$ to $g_1$ increases further.

\epsfxsize = 3.60in
\epsfysize = 2.80in

\newpage
\mbox{ }

\newpage
\section{Summary and conclusions}

In our previous work, we considered the unitary transformation of the spin-boson problem 
which resulted in a rotated Hamiltonian that was functionally simpler (although more
complicated mathematically).
One part of the rotated Hamiltonian ($\hat{H}_0$) seemed to give rise to an underlying unperturbed problem
in which the energy levels were very good, and in which no interaction was present between
the different levels.
Another part ($\hat{V}$) seemed to give a good account of the interaction responsible for the level splittings
at the anticrossings.
The remaining part ($\hat{W}$) is small in the large $n$ limit, and could be neglected.
In this work, we considered a generalization of the spin-boson model in which a three-level model
is coupled to an oscillator.
We were interested in whether a similar rotation could be developed, and whether it would exhibit 
similar useful properties.
It was conjectured in Ref. \cite{LarsonStenholm} that the rotation could be applied to multilevel
generalizations of the spin-boson problem, and the results presented here constitute such an example.

We succeeded in implementing such a rotation, which allowed us to develop analogous approximations
for energy levels and for level splittings at the anticrossings.
To study the approach, we developed a model test problem to provide a concrete example that could
be used to compare results from numerically exact calculations with those from the rotated version
of the problem.
This model problem itself illustrated interesting new effects not present in the spin-boson problem,
in which strong coupling between two levels result in one of the levels being pushed close to
the third level.
The resulting problem areas have no analog in the simpler spin-boson model.

The energy levels predicted from the rotated $\hat{H}_0$ problem are in excellent agreement with those
of the full $\hat{H}$ model away from the resonances, and away from the problem regions in the vicinity
of low-order resonances.
Approximate energy levels of the dressed problem were computed using the WKB approximation; which is
convenient for numerical calculations, and which is in good agreement with energy levels obtained through
brute force numerical solution of the $\hat{H}_0$ problem.
Energy levels were generally found to be good to better than 0.1 $\hbar \omega_0$ away from the problem
regions, which corresponds to a relative accuracy of better than 0.1 \% compared to the transition energies.
The accuracy is similar to what is observed in the spin-boson problem for similar transition energies and
dimensionless coupling strengths.

The computation of level splittings at the anticrossings requires the construction of interaction terms
in the dressed problem, which is more difficult than in the spin-boson problem because of complications
due to the three-level model.
We presented a straightforward method which relies on properties of the unitary transformation matrix,
and can be implemented in practice relatively simply using numerical differentiation.
Using these results, we found that the level splittings could be determined accurately 
using degenerate perturbation theory (similar to the case in the spin-boson problem);
as long as the dimensionless coupling strength ($g_1$ or $g_2$) of the transition in question was the
larger of the two, 
and as long as computations are done away from the problem area with low-order resonances on the other transition.
We expect that improved results for the level splittings can be obtained by employing
somewhat more sophisticated approximations than first-order degenerate perturbation
theory.
Few-state models in the rotated frame that take into account interfering resonances would
be expected to extend the range over which the level splittings could be approximated
accurately.

The dressed problem resulting from the unitary transformation provides a different view of
the coupled three-level and oscillator problem which allows us to understand the multiphoton
regime.
The new energy level approximation seems to work well, and may be useful for applications.
The approximation for calculating level splittings is less robust, but also seems to
be useful as long as applied in trouble-free regions.
We would expect this approach to apply to more complicated models involving more levels,
which would provide similar predictive capability for state energies.
In general, the approach should be most useful in regimes in which the oscillator energy
is small compared to all relevant transition energies.

\newpage

\appendix

\section{Solution of the eigenvalue equation}
\label{sec:cubic}

The characteristic equation for the energy eigenvalues that we obtained in the diagonalization
of the three-level system is the cubic equation 

$$
(E_1-E)(E_2-E)(E_3-E) - (E_1-E)2V^2y^2 - (E_3-E) 2 U^2 y^2 ~=~ 0
$$

\noindent
In this Appendix, we review the solution for the three energy eigenvalues (the solutions of which
are well known, but not usually presented in the form we discuss).
Our strategy will be to reduce it to the form of the triple angle sine formula

\begin{equation}
4 \sin^3 {\theta \over 3} - 3 \sin {\theta \over 3} ~=~ - \sin \theta
\end{equation}

\subsection{Offset}

We first define a new energy variable that is offset by the average energy of the three states

\begin{equation}
\epsilon ~=~ E - {E_1+E_2+E_3 \over 3}
\end{equation}

\noindent
which allows us to rewrite the characteristic equation in the form

\begin{equation}
\epsilon^3 - \alpha \epsilon ~=~ \beta
\end{equation}

\noindent
where $\alpha$ and $\beta$ are given by

{\small

\begin{equation}
\alpha ~=~ {E_1^2+E_2^2+E_3^2 - E_1E_2 - E_1E_3 - E_2E_3 \over 3} + 2U^2y^2 + 2 V^2 y^2
\end{equation}
}

{\small

$$
\beta 
~=~ 
{2 \over 27} (E_1^3+E_2^3+E_3^3)
-
{1 \over 9} [E_1^2(E_2+E_3)+E_2^2(E_1+E_3)+E_3^2(E_1+E_2)]
$$
\begin{equation}
+
{4 \over 9} E_1E_2E_3
+
{2 \over 3} U^2y^2(E_1+E_2-2E_3)
+
{2 \over 3} V^2y^2(E_2+E_3-2E_1)
\end{equation}

}

\subsection{Scaling}

Next, we scale according to

\begin{equation}
\epsilon ~=~ A \sin {\theta \over 3}
\end{equation}

\noindent
This leads to

\begin{equation}
4 \sin^3 {\theta \over 3} - {4 \alpha \over A^2} \sin {\theta \over 3} ~=~ {4 \beta \over A^3}
\end{equation}

\noindent
The triple angle sine formula is recovered with the identifications

\begin{equation}
{4 \alpha \over A^2} ~=~ 3
\end{equation}

\begin{equation}
{4 \beta \over A^3} ~=~ -\sin \theta
\end{equation}

\subsection{Energy eigenvalues}

Since the sine function is invariant under shifts of multiples of $2 \pi$

\begin{equation}
\sin \theta ~=~ \sin (\theta + 2 \pi) ~=~ \sin (\theta + 4 \pi)
\end{equation}

\noindent
there are three solutions to the triple angle sin formula

$$\sin \left ( {\theta \over 3} \right ), 
~ \sin \left ( {\theta + 2 \pi \over 3} \right ), 
~ \sin \left ( { \theta + 4 \pi \over 3 } \right )
$$

\noindent
Consequently, we obtain three solutions to the eigenvalue equation which we may write as

\begin{equation}
E 
~=~
{E_1 + E_2 + E_3 \over 3} 
+ 
A \sin \left ( {\theta \over 3} \right )
\end{equation}
\begin{equation}
E ~=~
{E_1 + E_2 + E_3 \over 3} 
+ 
A \sin \left ( {\theta+2\pi \over 3} \right )
\end{equation}
\begin{equation}
E ~=~ {E_1 + E_2 + E_3 \over 3} 
+ 
A \sin \left ( {\theta+4\pi \over 3} \right )
\end{equation}

\noindent
with

\begin{equation}
A ~=~ \sqrt{4 \alpha \over 3}
\end{equation}

\begin{equation}
\theta ~=~ \arcsin \left ( -{4 \beta \over A^3} \right )
\end{equation}

\noindent
The first of these energy expressions evaluates to the middle value of $E_1$, $E_2$, and $E_3$ when
$y=0$.  When $y=0$, the second evaluates to the maximum of $E_1$, $E_2$, and $E_3$, and the third
expression evaluates to the minimum of the three unperturbed energies.

\subsection{Algebraic expressions are inconvenient}

It is possible to develop algebraic expressions by taking advantage of
the addition formula

\begin{equation}
\sin(a+b) ~=~ \sin a \cos b + \cos a \sin b
\end{equation}

\noindent
The energy eigenvalues then can be written as

\begin{equation}
E 
~=~
{E_1 + E_2 + E_3 \over 3} 
+ 
A \sin \left ( {\theta \over 3} \right )
\end{equation}
\begin{equation}
E ~=~
{E_1 + E_2 + E_3 \over 3} 
+ 
A \left [ -{1 \over 2} \sin \left ( {\theta \over 3} \right ) + {\sqrt{3} \over 2} \cos \left ( {\theta \over 3} \right ) \right ]
\end{equation}
\begin{equation}
E ~=~ {E_1 + E_2 + E_3 \over 3} 
+ 
A \left [ -{1 \over 2} \sin \left ( {\theta \over 3} \right ) - {\sqrt{3} \over 2} \cos \left ( {\theta \over 3} \right ) \right ]
\end{equation}

To proceed, we require explicit expressions for $\sin (\theta/3)$ and $\cos (\theta/3)$.
We begin by writing for $\sin \theta$

\begin{equation}
\sin \theta 
~=~ 
{e^{i \theta} - e^{-i \theta } \over 2i}
~=~
- {3^{3 \over 2} \beta \over 2 \alpha^{3 \over 2}}
\end{equation}

\noindent
We can solve for $e^{i \theta}$ to obtain

\begin{equation}
e^{i \theta} 
~=~
-i {3^{3 \over 2} \beta \over 2 \alpha^{3 \over 2}}
~\pm~
\sqrt{1 -  {27 \beta^2 \over 4 \alpha^3} }
\end{equation}

\noindent
Either choice of sign is acceptable, but we will choose a $+$ sign for what follows.
This allows us to write

{\small

\begin{equation}
\sin \left ( {\theta \over 3} \right ) 
~=~
{1 \over 2i}
\left \lbrace
\left [
\sqrt{1 -  {27 \beta^2 \over 4 \alpha^3} }
-i {3^{3 \over 2} \beta \over 2 \alpha^{3 \over 2}}
\right ]^{1 \over 3}
-
\left [
\sqrt{1 -  {27 \beta^2 \over 4 \alpha^3} }
-i {3^{3 \over 2} \beta \over 2 \alpha^{3 \over 2}}
\right ]^{-{1 \over 3}}
\right \rbrace
\end{equation}

}

{\small

\begin{equation}
\cos \left ( {\theta \over 3} \right ) 
~=~
{1 \over 2}
\left \lbrace
\left [
\sqrt{1 -  {27 \beta^2 \over 4 \alpha^3} }
-i {3^{3 \over 2} \beta \over 2 \alpha^{3 \over 2}}
\right ]^{1 \over 3}
+
\left [
\sqrt{1 -  {27 \beta^2 \over 4 \alpha^3} }
-i {3^{3 \over 2} \beta \over 2 \alpha^{3 \over 2}}
\right ]^{-{1 \over 3}}
\right \rbrace
\end{equation}

}

\noindent
Unfortunately, these expressions are complicated in a way that makes them inconvenient
for calculations.

\subsection{Analytic WKB expression}

It is possible to use these results to develop analytic expressions for the WKB
approximation for the energy levels discussed in Section 3.
In the case of the middle level, we may write the approximation as

\begin{equation}
E_2(g_1,g_2)
~=~
{1 \over \pi} \int_{-\epsilon}^\epsilon {E_2(y) \over \sqrt{\epsilon - y^2}} dy
\end{equation}

\noindent
After substituting in for $E_2(y)$ we obtain

$$
E_2(g_1,g_2)
~=~
{E_1 + E_2 + E_3 \over 3} 
\ \ \ \ \ \ \ \ \ \ \ \ \ \ \ \ \ \ \ \ \ \ \ \ \ \ \ 
\ \ \ \ \ \ \ \ \ \ \ \ \ \ \ \ \ \ \ \ \ \ \ \ \ \ \ 
\ \ \ \ \ \ \ \ \ \ \ \ \ \ \ \ \ \ \ \ \ \ \ \ \ \ \ 
$$
\begin{equation}
+
{1 \over \pi} 
\int_{-\epsilon}^\epsilon 
{A(y) \sin \left ( \displaystyle{1 \over 3} \arcsin \left [ -\displaystyle{4 \beta(y) \over A^3(y)} \right ] \right )  \over \sqrt{\epsilon - y^2}} dy
\end{equation}

\noindent
Similar analytic expressions can be written for the other two levels directly.

\end{document}